\documentclass[fleqn,12pt,twoside]{article}
\usepackage{espcrc1}

\usepackage{graphicx}
\usepackage{epsfig}
\usepackage[figuresright]{rotating}

\newcommand{\AmS}{{\protect\the\textfont2
  A\kern-.1667em\lower.5ex\hbox{M}\kern-.125emS}}
\newcommand{\be}{\begin{equation}}
\newcommand{\ee}{\end{equation}}
\newcommand{\bea}{\begin{eqnarray}}
\newcommand{\eea}{\end{eqnarray}}
\newcommand{\f}{\frac}

\newcommand{\Ga}{\Gamma}
\newcommand{\ga}{\gamma}
\newcommand{\bra}{\langle}
\newcommand{\ket}{\rangle}

\newcommand{\Seff}{S_{\rm{eff}}}

\title{Resonances and the thermonuclear reaction 
rate\thanks{Supported by FAPESP.}}
\author{M.~S.~Hussein\address[USP]
{Nuclear Theory and Elementary Particle Phenomenology 
Group, Instituto de F\'\i sica, Universidade de S\~{a}o Paulo,
Caixa Postal 66318, 05315-970 S\~{a}o Paulo, SP, Brazil},
M.~Ueda\address[ANCT]{Akita National College of Technology, 
Iijima Bunkyo-cho 1-1, Akita, 011-8511, Japan}, 
A.~J.~Sargeant\addressmark[USP]
and
M.~P.~Pato\addressmark[USP]}

\begin{document}
\maketitle

\section{Abstract}

We present an approximate analytic expression for thermonuclear reaction rate 
of charged particles when the cross section
contains a single narrow or wide resonance described by a Breit-Wigner shape.
The resulting expression is uniformly valid as the effective energy and 
resonance energy coalesce. We use our expressions to calculate the reaction 
rate for $^{12}$C(p,$\ga$)$^{13}$N.
\section{Introduction}\label{intro}
Evaluations of thermonuclear reaction rates require the folding
of nuclear cross sections with the Maxwell--Boltzmann
distribution.
In recent compilations of such rates for charged particles
\cite{Angulo:1999} ($Z$=1--14), \cite{Iliadis:2001} ($A$=20--40) a combination
of numerical integration and analytic techniques appropriate to the 
energy dependence of the cross section for each reaction pair 
were used. A review of the analytical techniques of nuclear reaction rate 
theory can be found in Ref.~\cite{Mathai:2002}.
Ref. \cite{Nelson:2000} has obtained results different to those obtained in
Ref. \cite{Angulo:1999} due to the inclusion of resonances in their numerical 
integrations. Ref. \cite{Angulo:1999} calculated the contribution of 
narrow resonances in the simplest possible 
manner which is to approximate the Maxwell--Boltzmann distribution by 
its value at the resonance energy. This approximation can be expected to be good
for resonances which do not overlap significantly with the Gamow peak.

In Ref. \cite{Ueda:2000xa} we developed an asymptotic expansion for the
thermonuclear reaction rate in terms of the
effective astrophysical $S$-factor, $\Seff$, using the method of Dingle
\cite{Dingle:1973}. The method  may be used in cases where $S(E)$ can be reliably 
expanded as a Taylor series. Two alternative expressions $\Seff$
where obtained by expanding $S(E)$ about $E$=$0$ and about $E$=$E_0$, 
$E_0$ being the effective energy of the Gamow window. From these expressions,
all the approximate formulae for $\Seff$ commonly used \cite{Adelberger:1998qm}
when $S(E)$ is slowly varying may be obtained as special cases. 
The validity of the expressions derived in Ref. \cite{Ueda:2000xa} is limited
by the radii of convergence of the Taylor series expansions of $S(E)$, 
that is, by the location of the poles of $S(E)$. The poles may be due bound 
states of the composite nucleus. Illustrative is the case of 
$^7$Be + $p$ $\rightarrow$ $^8$B + $\ga$ \cite{Jennings:1998}
for which $S(E)$ has a pole at $E$=$-E_B$, $E_B$=$137.5$ keV 
being the binding energy of $^8$B.
(Ref.~\cite{Mukhamedzhanov:2002} discusses effects other than a
sub-threshold pole which can produce the low energy rise seen 
in the $S$-factors of several capture reactions.)
There may also be poles in $S(E)$ due
to resonances of the reaction pair in which case the cross section may be 
parameterised in the region of the resonance by the Breit-Wigner form 
(see Eq.~(\ref{bw}) below) which has two poles, or the Lorentzian form 
\cite{Goriely:1998} which has four poles. 

\begin{figure}[thb]
\centering
\includegraphics[width=.7\textwidth]{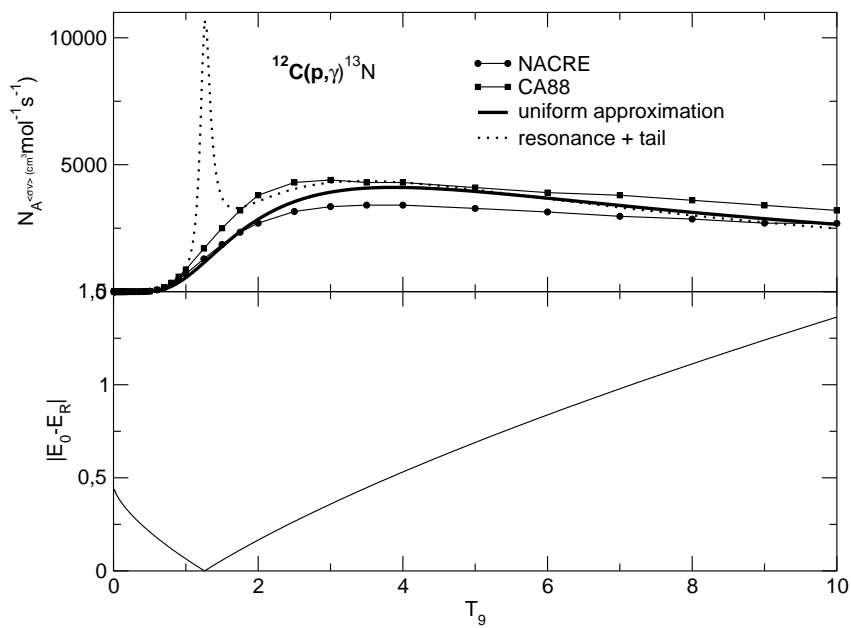}
\caption{\label{c12pg}  The reaction rate for 
$^{12}$C(p,$\ga$)$^{13}$N as function of $T_9$. The circles show the
rates from Ref.~\cite{Angulo:1999}, the squares Ref.~\cite{Caughlan:1988},
the solid lines the uniform approximation, Eq.~(\ref{unirate}), and the
dotted lines the resonance-plus-tail approximation, Eq.~(\ref{resandtail}).
We took the resonance parameters from Ref.~\cite{Rolfs:1974} 
($E_r$=457 keV and $\Ga_r \approx \Ga_p(E_r)$=39 keV
[both being laboratory frame values] and $\sigma(E_r)$=130 $\mu$ b).}
\end{figure}
For a given reaction cross section $\sigma(E)$, the Maxwellian averaged reaction 
rate per particle pair is given by \cite{Rolfs:1988}
\begin{equation}
\langle \sigma v \rangle = \left ( \frac{8}{\pi \mu} \right )^{1/2} 
\left ( \frac{1}{k_B T} \right )^{3/2} \int_0^{\infty} dE 
 E\sigma(E)e^{-E/k_BT} 
\label{rate}
\end{equation}
where is the reduced mass of the two collision partners, 
$k_B$ is the Boltzmann constant, and $T$ the environmental
temperature.
Here, we consider a $(p, \ga)$ resonant reaction for which the cross section 
is given by
\begin{equation}
\sigma(E)= \frac{\pi \hbar^2}{2 \mu} \frac{1}{E} 
\omega \frac{\Ga_p(E)\Ga_\ga}{(E-E_r)^2+\Ga_r^2/4} \ \ ,
\label{bw}
\end{equation}
where $\omega$ is the statistical spin factor, $\Ga_\ga$ is the $\ga$-decay width and
$\Ga_r$ is the total width at the resonance energy.
The energy dependence of the proton-decay width $\Ga_p(E)$
is determined by the Coulomb barrier for low enough energies and may be written as
$\Ga_p(E)$=$\ga_p \exp\left[-2 \pi \eta(E)\right]$,
with $\ga_p$ energy independent. The Sommerfeld parameter, $\eta(E)$, is given by
$2\pi\eta(E)$=$\sqrt{E_G/E}$, where 
$E_G$=$b^2$$\equiv$$2\mu c^2\left(\pi Z_1Z_2\alpha\right)^2$,
$\alpha$ being the fine structure constant.
The effective energy of the Gamow window at temperature $T$ is found to be
$E_0$=$\left[E_G^{1/2}k_BT/2\right]^{2/3}$.

Substituting Eq.~(\ref{bw}) into Eq.~(\ref{rate}) and 
introducing the dimensionless variables 
$x$=$E/k_BT$, $x_r$=$E_r/k_BT$ and $x_I$=$\Ga_r/2/k_BT$
one can rewrite the reaction rate as
\begin{eqnarray}
\langle \sigma v \rangle &=&
\left ( \frac{8}{\pi \mu} \right )^{1/2} 
\left ( \frac{1}{k_B T} \right )^{5/2} 
\frac{\pi \hbar^2\omega\ga_p\Ga_\ga}{2 \mu}
\f{1}{x_I}\Im\int_0^{\infty} dx \frac{e^{-F(x)}}{x-x_p}. \label{ex1}
\end{eqnarray}
where
$F(x)$=$x$+$\sqrt{x_G/x}$, $x_p$=$x_r$+$ix_I$ and $x_G$=$a^2$$\equiv$$E_G/k_BT$.
The function $F(x)$ has a single minimum in the the interval $[0,\infty)$ 
from which it increases monotonically to infinity as $x$ approaches zero and infinity.
The position of the minimum is given by
$x_0$=$E_0/k_BT$=$\left[x_G^{1/2}/2\right]^{2/3}$.
The principal contributions 
come from the neighbourhood of the critical points $x_0$ and $x_p$ 
\cite{Dingle:1973,Bleistein:1975}.
\section{Conventional approximation}
\label{conv}
When the resonance and the Gamow window are well separated the contribution from each
to the reaction rate can be estimated separately and the two contributions summed.
Let us write
$\bra\sigma v\ket\approx \bra\sigma v\ket_r+\bra\sigma v\ket_G$,
where $\bra\sigma v\ket_r$ is the contribution from the resonance and $\bra\sigma v\ket_G$ 
the contribution from the Gamow window. 
The resonance contribution may be estimated \cite{Rolfs:1988}
by approximating the exponential in Eq.~(\ref{ex1}) by it's value at the resonance energy, $x_r$,
so that $\int_0^{\infty}dxe^{-F(x)}/[(x-x_r)^2 + x_I^2]
\approx e^{-x_r-\sqrt{x_G/x_r}}\left(\pi/2+\theta_r\right)/x_I$,
where $\theta_r$=$\tan^{-1}\left(x_r/x_I\right)$.

Similarly, the Gamow window contribution is estimated by approximating the Breit-Wigner
factor by it's value at the location of the Gamow peak, $x_0$, so that
$\int_0^{\infty}dxe^{-F(x)}/[(x-x_r)^2 + x_I^2]
\approx\frac{2}{3}\sqrt{\pi \tau}e^{-\tau}/[(x_0-x_r)^2 + x_I^2]$.
We have also used the Gaussian approximation 
$F(x)$$-$$F(x_0)$$\approx$$\left(x-x_0\right)^2/\left(\Delta_x/2\right)^2$
\cite{Rolfs:1988} where
$\Delta_x$=$\sqrt{8/F''(x_0)}$=$4\sqrt{x_0/3}$ is
the effective width of the Gamow window in units of $k_BT$ and
$\tau$=$F(x_0)$=$3x_0$.

On thus obtains
\be\label{resandtail}
\langle \sigma v \rangle\approx\left(2/\mu\right)^{1/2}
\f{1}{(k_BT)^{3/2}}
\left[\pi^{1/2}\Ga S(E_r)e^{-E_r/k_BT-\sqrt{E_G/E_r}}
+\Delta S(E_0)e^{-\tau}\right],
\ee
where we have introduced the astrophysical $S$-factor, 
$S(E)$=$\sigma(E)Ee^{\sqrt{E_G/E}}$
and the width of the Gamow window $\Delta$=$4(E_0k_B T/3)^{1/2}$.
The second term of Eq.~(\ref{resandtail}) is the zeroth order term of an asymptotic expansion of $I_G$ 
in powers of $1/\tau$ \cite{Ueda:2000xa}
\section{Uniform approximation}
\label{uni}
We introduce a new integration variable $t$ implicitly through $F(x)$$-$$F(x_0)$=$t^2$
and rewrite the reaction rate in terms of the function 
$\Phi(t)$=$\frac{dx}{dt}\frac{t-t_p}{x-x_p}$.
The position of the pole in $t$-space is then given by
$t_p$=$\left[x_p+\sqrt{x_G/x_p}-\tau\right]^{1/2}$.
In order to obtain an expansion which is uniform as $x_p$ is made to 
approach or coincide with $x_0$,
$\Phi(t)$ is expanded as follows \cite{Bleistein:1975,Chester:1957,Berry:1966}: 
$\Phi(t)$=$\sum_{n=0}^{\infty}\left[\alpha_n+(t-t_p)\beta_n\right]\left[t(t-t_p)\right]^n$.
This expansion has the property that
it gives the value of the function $\Phi(t)$ and all it's derivatives 
exactly at both $t=0$ and $t=t_p$.
Let us approximate $\Phi(t)$ by the truncation
$\Phi_0(t)$=$\alpha_0$+$(t-t_p)\beta_0$.
The expansion coefficients are found to be $\alpha_0$=$1$ and 
$\beta_0$=$1/t_p$+$2\sqrt\tau/3/(x_0-x_p)$.
Introducing the complementary error function  
\cite{Abramowitz:1970} we obtain
\be\label{unirate}
\langle \sigma v \rangle\approx\left(2/\mu\right)^{1/2}
\f{e^{-\tau}}{(k_BT)^{3/2}}
\left[\Ga S(E_r)
\Im\left\{i\sqrt{\pi}e^{-t_p^2}{\rm erfc}(-i t_p)
+\frac{1}{t_p}\right\}
+\Delta S(E_0)\right].
\ee
\section{Reaction rate for $^{12}$C(p,$\ga$)$^{13}$N}
For the $S$-factor at the resonance energy we use
$S(E_r)$=$\f{2\pi\hbar^2}{\mu}\f{(\omega\ga)_r}{\Ga}e^{\sqrt{E_G/E_r}}$
and at the effective energy
$S(E_0)$=$S(E_r)\f{\Ga^2/4}{(E_0-E_r)^2+\Ga^2/4}$,
where the resonance strength is given by
$(\omega\ga)_r$=$\f{\sigma(E_r)\mu\Ga_r E_r}{2\pi\hbar^2}$.
We have calculated the reaction rate for $^{12}$C(p,$\ga$)$^{13}$N using
Eqs.~(\ref{unirate}) and (\ref{resandtail}) including only the lowest
resonance and the results are displayed in Figure~\ref{c12pg}.
The exact integral is not shown
as it is indistinguishable from the uniform approximation for this case.
It is apparent
that the resonance-plus-tail approximation cannot be used at temperatures for which
$E_r$$\sim$$E_0$. For comparison we have also shown the rates of
Ref.~\cite{Caughlan:1988} which used a different analytic approach and
those of Ref.~\cite{Angulo:1999} which also included the contribution of the
second resonance at 1698 keV.
\bibliography{astro,sargeant,books,asymp}

\begin{thebibliography}{10}

\bibitem{Angulo:1999}
C.~{Angulo}, M.~{Arnould}, M.~{Rayet} \emph{et~al.}, Nucl. Phys. \textbf{A656}
  (1999) 3.

\bibitem{Iliadis:2001}
C.~{Iliadis}, J.~M. {D'Auria}, S.~{Starrfield} \emph{et~al.}, Astrophys. J.
  Suppl. \textbf{134} (2001) 151.

\bibitem{Mathai:2002}
A.~M. {Mathai} and H.~J. {Haubold}, Astrophys. Space Sci. \textbf{282} (2002)
  265.

\bibitem{Nelson:2000}
S.~O. {Nelson}, E.~A. {Wulf}, J.~H. {Kelley} \emph{et~al.}, Nucl. Phys.
  \textbf{A679} (2000) 199.

\bibitem{Ueda:2000xa}
M.~Ueda, A.~J. Sargeant, M.~P. Pato \emph{et~al.}, Phys. Rev. C \textbf{61}
  (2000) 045801.

\bibitem{Dingle:1973}
R.~B. Dingle, \emph{Asymptotic expansions: their derivation and interpretation}
  (Academic Press, New York \& London, 1973).

\bibitem{Adelberger:1998qm}
E.~G. {Adelberger}, S.~M. {Austin}, J.~N. {Bahcall} \emph{et~al.}, Rev. Mod.
  Phys. \textbf{70} (1998) 1265.

\bibitem{Jennings:1998}
B.~K. {Jennings}, S.~{Karataglidis} and T.~D. {Shoppa}, Phys.Rev. C \textbf{58}
  (1998) 3711.

\bibitem{Mukhamedzhanov:2002}
A.~M. {Mukhamedzhanov} and F.~M. {Nunes}, Nucl. Phys. \textbf{A708} (2002) 437.

\bibitem{Goriely:1998}
S.~{Goriely}, Phys. Lett. B \textbf{436} (1998) 10.

\bibitem{Caughlan:1988}
G.~R. {Caughlan} and W.~A. {Fowler}, At. Data and Nucl. Data Tables \textbf{40}
  (1988) 283.

\bibitem{Rolfs:1974}
C.~{Rolfs} and R.~E. {Azuma}, Nucl. Phys. \textbf{A227} (1974) 291.

\bibitem{Rolfs:1988}
C.~E. Rolfs and W.~S. Rodney, \emph{Cauldrons in the Cosmos} (Chicago Univ.
  Press, 1988).

\bibitem{Bleistein:1975}
N.~Bleistein and R.~A. Handelsman, \emph{Asymptotic Expansions of Integrals}
  (Holt, Rinehart and Winston, New York (Reprinted in 1986 by Dover, New
  York.), 1975).

\bibitem{Chester:1957}
C.~Chester, B.~Friedman and F.~Ursell, Proc. Camb. Phil. Soc. \textbf{53}
  (1957) 599.

\bibitem{Berry:1966}
M.~V. Berry, Proc. Phys. Soc. \textbf{89} (1966) 479.

\bibitem{Abramowitz:1970}
M.~Abramowitz and I.~A. Stegun (eds.) \emph{Handbook of Mathematical Functions}
  (Dover, 1970).

\end{thebibliography}
\bibliographystyle{npa}
\end{document}